\documentclass[doublecol]{epl2}

\usepackage{amsmath, amssymb, graphics, graphicx}
\usepackage{textcomp, subfigure}
\usepackage{color}
\usepackage[final]{pdfpages}

\usepackage{sistyle}					

\newcommand{\eps}{\epsilon}

\definecolor{linkcolor}{rgb}{0,0,0.6} 
\usepackage[pdftex,colorlinks=true, pdfstartview=FitV, linkcolor= linkcolor, citecolor= linkcolor, urlcolor= linkcolor, hyperindex=true,hyperfigures=true]{hyperref} 

\begin{document}
\title{\bf   On  the transient Fluctuation Dissipation Theorem  \\
              after a quench at a critical point}
\author{   Isaac Theurkauff, Aude Caussarieu, Artyom Petrosyan, Sergio Ciliberto }
\institute{ Universit\'e de Lyon Laboratoire de Physique,  \'Ecole Normale Sup\'erieure,  C.N.R.S. UMR5672 
\\ 46 All\'ee d'Italie, 69364 Lyon, France \\}

\begin{abstract}
{The  Modified Fluctuation Dissipation Theorem (MFDT) proposed by G. Verley et al.
{\it (EPL 93, 10002, (2011))} for non equilibrium transient states is experimentally studied.  We apply MFDT 
 to the  transient relaxation dynamics of the director of a liquid crystal after a quench close to the critical point of the Fr\'eedericksz transition (Ftr), which has several  properties of a second order phase transition driven by an electric field. 
Although  the standard Fluctuation Dissipation Theorem (FDT)  is not satisfied, because the system is strongly out of equilibrium, the MFDT is perfectly verified during the transient in a system which is only partially described by Landau-Ginzburg (LG) equation, to which our observation are compared. The results can be useful in the study of  material aging. }
\end{abstract}
\maketitle




After a  sudden change of a thermodynamic parameter, such as temperature, volume and  pressure, several systems and materials may present an extremely slow relaxation  towards equilibrium. During this slow relaxation, usually called aging, these systems remain out-of equilibrium for a very long time, their properties are slowly evolving and  equilibrium relations are not necessarily satisfied during aging. 
Typical and widely studied examples of this phenomenon are glasses and colloids where many questions on their relaxation dynamics still remain open \cite{Cavagna,Vincent}. Thus in order to understand the minimal ingredients for aging, slow relaxations have been studied theoretically in second order phase transitions when the system is rapidly quenched from an initial value of the control parameter to the critical point \cite{Berthier, Holdsworth,Godreche,Gambassi}. Because of the critical slowing down and the divergency of the  correlation length the relaxation dynamics of the critical model  shares several features of the aging of more complex materials. One of the questions analyzed in this models is the validity of the Fluctuation Dissipation Theorem (FDT) during the out of equilibrium relaxation\cite{Peliti,Hanggi,Hanggi2011,Puglisi}. In equilibrium,  FDT imposes a relationship between  the response of the system to a small external perturbation  and  the correlation of the spontaneous thermal fluctuations. When the system is out of equilibrium FDT does not  necessarily hold and it has been generalized as 
\begin{eqnarray}
k_B T \ X(t,t_w) \ \chi(t,t_w)&=& C(t,t) - C(t,t_w)
\label{eq_GFDT}
\end{eqnarray}
where $k_B$ is the Boltzmann constant, $T$ the bath temperature, $C(t,t_w)=<O(t)O(t_w)>-<O(t)><O(t_w)>$ ($<.>$ stands for ensemble-average) the correlation function of the observable $O(t)$.
The function $\chi(t,t_w)$ is the response  to a small step perturbation, of  the conjugated variable $h$ of $O$,  applied at time $t_w < t$  : 
\begin{eqnarray} \chi(t,t_w)={< O(t)>_{h} - < O(t)>_0 \over h}\vert _{h\rightarrow 0}
\label{eq_chi}
\end{eqnarray}
where $<O(t)>_{h}$ and $<O(t)>_{o}$ denote respectively the mean perturbed and unperturbed time evolution.
The function $X(t,t_w)$ is  equal 1 in equilibrium whereas in out-equilibrium it measures the amount of the FDT violation and it has been used in some cases to define an effective temperature $T_{eff}(t,t_w)= X(t,t_w) T$.  

The above mentioned models of the quench at critical points allows a precise analysis of the pertinence of this definition of $T_{eff}$\cite{Peliti}. In spite of the theoretical interest of these models only one experiment has been performed on the slow relaxation dynamics after a quench at the critical point \cite{Joubaud_PRL}. The role of this letter is to experimentally analyze the theoretically predictions in a real system affected by finite size effects and unavoidable imperfections.  We also analyze another important aspects  of the FDT in out of equilibrium systems. Indeed several generalizations  of FDT has been proposed \cite{Hanggi,Hanggi2011,Speck,Chetrite,Puglisi,Harada,Lipiello,Maes,Prost}
 but almost all of them can be applied to non-equilibrium steady states \cite{Gomez,Blicke} and are not useful for the transient time evolution which follows the quench at the critical point. As far as we know there are only two formulations of FDT \cite{Lacoste_EPL,Baiesi_2014}, which are useful for these transient states and the second purpose of this letter is  to discuss  the application of the Modified FDT (MFDT) of ref.\cite{Lacoste_EPL}.

Before describing the experimental set-up we summarize briefly the formulation of the MFDT of ref.\cite{Lacoste_EPL} . Let us consider  the relaxation dynamics of  a system, which has been submitted at time  $t=0$ to a sudden change of its control parameters.   At time $t$ after the quench, this relaxation is characterized by the variable $x(t)$, by the observable $O(x(t))$  and by the probability density function  $\pi(x(t),h(t))$. Here $h(t)$ is an external control parameter which is used to perturb the dynamics.   We  define 
a pseudopotential 
$\Psi(t,h)=-\ln [\pi(x(t),h)] $ 
and an  observable $B(t)={-\partial_h \psi(t,h) }\vert _{h\rightarrow 0}$, 
with $h\ne0$ constant   for $t>0$ and  $h=0$ for $t<0$. 
The  MFDT reads:
\begin{eqnarray}
\chi(t,t_w)&=&<B(t) O(t)> - <B(t_w)O(t)> 
\label{eq_MFDTa}
\end{eqnarray}
Eq.\ref{eq_MFDTa} defines the response function $\chi(t,t_w)$ of $O(t)$ to a step perturbation of $h$ applied at $t_w$, with $0<t_w<t$. Notice that in this case $h$ can be any parameter of the system and it does not {need} to be the conjugated variable of $O(t)$.  
In this letter we will analyze how MFDT (eq.\ref{eq_MFDTa}) can be applied to the experimental data using   the quench at the critical point in a non ideal system.  
\begin{figure}[h]  
   \includegraphics[width=0.49\textwidth]{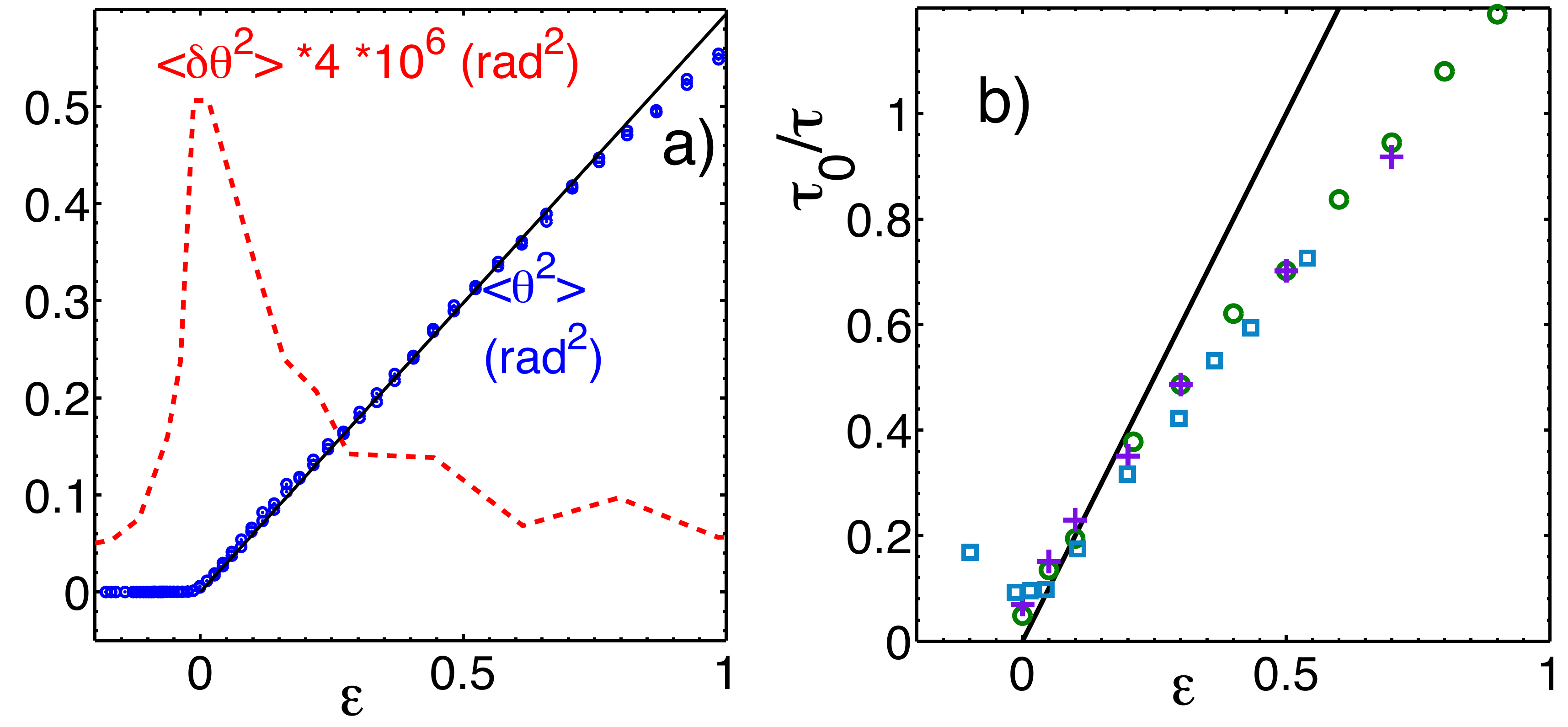}    
     \caption{Phase diagram of the Freedericksz Transition in 5CB. a) Dependence on $\epsilon$ of $\theta_o^2=<\theta>^2$ (blue dots)  and of the variance ( dashed red line) of $\theta$ multiplied by $4 \ 10^6$ to be on the same scale. The solution of the LG equation ($\mu=0$)  is the straight black line. b) $\tau_o/\tau$ is plotted as a function of $\epsilon$. The prediction of LG ($\mu=0$)  is the black straight line. The measured  relaxation time deviates significantly from the predicted one even for values of $\epsilon$ where the stationary solution of LG, plotted in a) seems to reproduce the data. }
 \label{fig_phase_diagram} 
\end{figure}

The experimental system where we study these properties  and the MFDT is the Freedericksz Transition (FrTr) in a nematic  liquid crystal (LC)   submitted  to an external electric field $\vec{E}$. Specificlly in our experiment we use the 5CB (p-pentyl-cyanobiphenyl, 5CB, produced by Merck). The experimental apparatus has been already described  \cite{Caussarieu_APL,Caussarieu_EPL} and we summarize here only the main features. 
 The LC is confined between two glass plates, separated by a distance $L =${13.5}$\,\mu\mbox{m}$.
The surfaces in contact with LC molecules are coated by ITO to apply an electrical field. Then, a polymer layer (rubbed PVA) is deposited to insure a 
strong anchoring of the 5CB molecules in a direction parallel to the plates. In the absence of any external field, the molecules in the cell align
parallel to those anchored at the surfaces. Applying a voltage difference $U$ between the electrodes, the liquid crystal is submitted
to an electrical field perpendicular to the plates. To avoid polarization, the applied voltage is modulated at a frequency f = 10 kHz 
$\left[U = \sqrt{2}U_0\mbox{cos}(2\pi f t)\right]$. When $U_0$ exceeds a threshold value $U_c$, the planar states becomes unstable and the molecules rotate  to align with the electrical field.  To quantify the transition we measure the spatially averaged 
alignment of the molecules determined by the angle $\theta$ between the molecule director and the surface.  Such measurement relies upon the anisotropic properties of the nematic. This optical anisotropy can be precisely measured using a very sensitive polarization interferometer \cite{Caussarieu_APL} which gives a signal $\varphi\propto \theta^2$.
At $U\simeq U_c$ the dynamics is usually described by a Landau-Ginzburg (LG) equation although as pointed out in ref.\cite{Caussarieu_EPL} this is a very crude approximation, which has several drawbacks. In a very first approximation 
 the dynamics of the mean relaxation $\theta_0(t) = \left\langle \theta(t) \right\rangle$ 
is ruled by the following Ginzburg-Landau equation :
\begin{equation}
\tau_0 \dot{\theta}_0 =\eps \theta_0 -\frac{\alpha}{2}\left( \theta_0^3 - \mu^3 \right) 
\label{eq_landau}
\end{equation}
where $\eps=(U^2-U_c^2)/U_c^2$ is the reduced control parameter,  the $\tau_o$ is a characteristic time of the LC  and $\alpha$ a parameter which depends on the elastic and electric anisotropy of the LC. For 5CB $\alpha=3.36$ and $\tau_o=2.4s$  for the cell thickness $L=13.5\mu m$. 
 The residual angle $\mu\simeq 0.1$ at $\eps=0$  comes from cell assembling and preparation and has been discussed in ref.\cite{Caussarieu_EPL}.
 Furthermore $\tau_o$ is not strictly constant but it slightly depends on $\theta_o^2$.  
The dimensional equation for the fluctuations $\delta \theta(t)= \theta(t)-\theta_0(t)$  is 
\begin{eqnarray}
\gamma A L \dot {\delta \theta} &=& K [  (\eps-{3\over 2} \alpha \theta_o^2)  \   \delta \theta + \delta \eps \ \theta_o ]  + \eta  \label{eq_delta}\\
\text{ with } \  \  \  K&= & {  \pi^2 k_1 A \over L}
\label{eq_fluct}
\end{eqnarray}
where $A$ is the laser cross section, $k_1$ is one of the elastic constant of LC  and $\eta$ is a delta correlated thermal noise such that 
$<\eta(t)\eta(t')> =k_B T (\gamma A L) \delta (t-t')$. The term with $\delta \eps $   takes into account that  during the measure of  the response  we  have  to  perturb the value of $\epsilon$ by applying  a short pulse of duration $\tau_p$ and amplitude $\delta \epsilon$.
The two eqs.\ref{eq_landau},\ref{eq_delta} describe in principle the dynamics of the mean deflection $\theta_o$ and of the fluctuations $ {\delta \theta}$.
{ However there are several discrepancies with the experimental data which are  widely discussed in ref.\cite{Caussarieu_EPL}.  We summarize here the most important, which are useful for the discussion.   The phase diagram of FrTr in 5CB and the relaxation times are plotted in figs. \ref{fig_phase_diagram} a),b) respectively.  The solution  of LG $\theta_o^2\simeq 2\epsilon/\alpha$ ($\mu=0$ in eq.\ref{eq_landau}) reproduces the stationary  experimental data for $\epsilon \le 0.6 $.  Instead the measured  relaxation time
 (fig. \ref{fig_phase_diagram} b) deviates significantly from the predicted one even for values of $\epsilon$ where the stationary solution of LG,   (fig. \ref{fig_phase_diagram} a) seems to reproduce the data. The characteristic time  increases 
 but  it does not diverge because of $\mu\ne 0$ (see fig.\ref{fig_phase_diagram} and ref.\cite{Caussarieu_EPL}). In fig. \ref{fig_phase_diagram} a) the variance $\sigma_\theta^2$ of $\delta$ is plotted too.  From eq.\ref{eq_delta} this variance is $\sigma_\theta^2=k_BT/ ( K (\eps-{3\over 2} \alpha \theta_o^2))$ which does not diverge at $\epsilon=0$ because $\mu\ne 0$. \\
Thus although eq. \ref{eq_landau} and eq.\ref{eq_delta} are  only a rough approximation of the FrTr dynamics, especially at $\epsilon>0.1$, we use them to fix the framework, and because they are  very close to the theoretical mean field approach to the quench at critical point discussed in ref. \cite{Gambassi}.   Thus it is interesting to check the analogies and differences with respect to the general theory. }

\begin{figure}[h]  
   \includegraphics[width=0.49\textwidth]{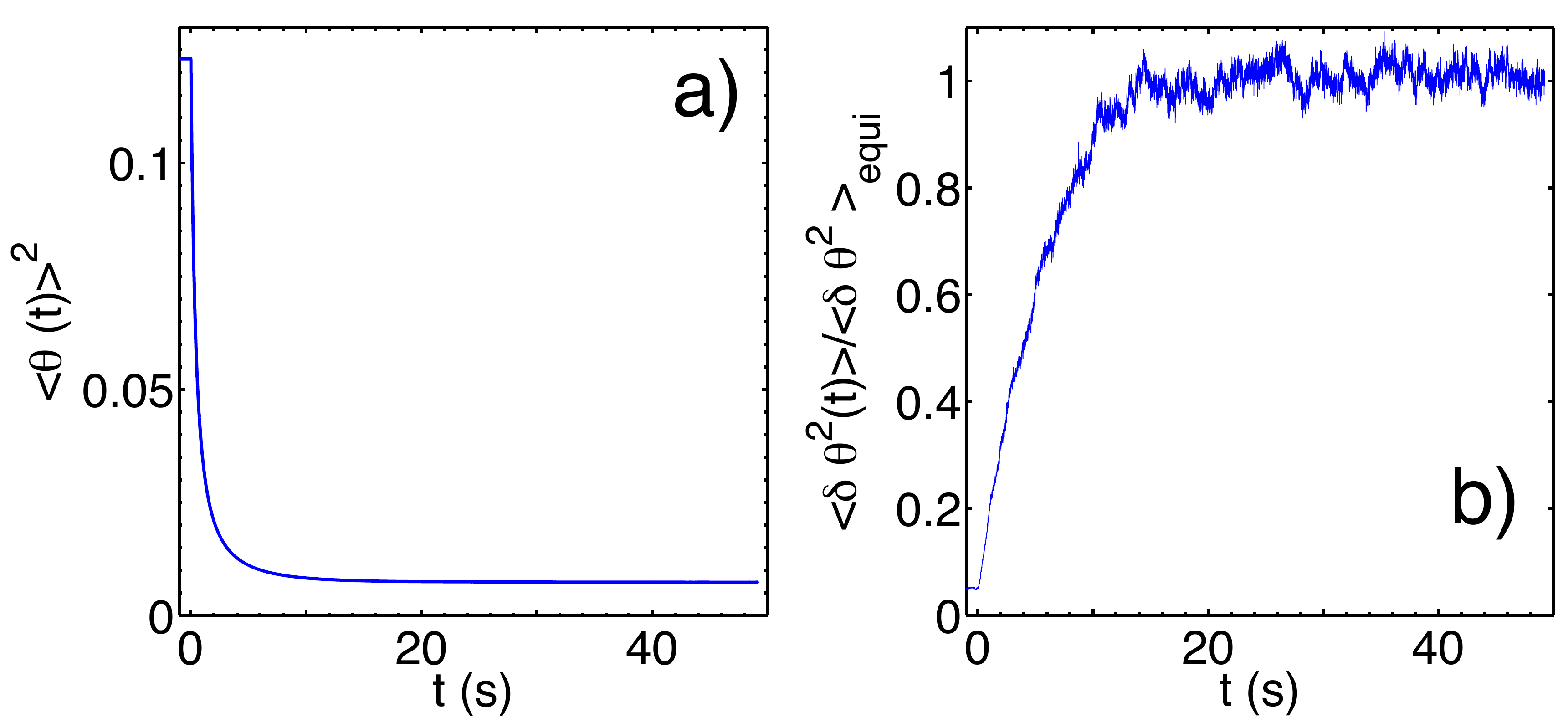}    
     \caption{Quench close to the critical point from $\epsilon_i=0.25$ to $\epsilon_f=0.01$. a) Time evolution of the order parameter $\theta_o^2$ before and after the quench preformed at $t=0$. b) Time evolution of the variance as a function of time. The variance and the mean $\theta^2(t)$ has been obtained by performing  $3000$ quenches and then making an ensemble average  on the quenches at each time. 
}
 \label{fig_thet_var} 
\end{figure}

The quench is performed  by commuting $\epsilon$ from an initial value $\epsilon_i$ to an $\epsilon_f \simeq 0$ at $t=0$. As an example we show in fig.\ref{fig_thet_var} the time evolution of $\theta_o$ for a quench from $\epsilon_i=0.25$ to $\epsilon_f=0.01$. The system is relaxing from its initial equilibrium value towards the new one. We describe here  the time evolution of the statistical properties and we will discuss at the end the dependence on the initial and final $\epsilon$ values. The mean values of the statistical properties are obtained by repeating the quench at least $3000$ times.  

The time evolutions of  $\theta_0^2(t)$ and of the variance $\sigma_\theta^2(t)$    are shown in figs.\ref{fig_thet_var}a) and b).  We see that both quantities relax from the initial to the final equilibrium values ,which are $\theta_e^2\simeq 2\epsilon/\alpha$ 
and $\sigma_\theta^2(t)=k_BT/ ( K (\eps-{3\over 2} \alpha \theta_e^2))$, where  
$\theta_e^2$ is the equilibrium value, which is not exactly $2\epsilon/\alpha$ because of the presence of the imperfect bifurcation $\mu\ne 0$ (see fig.\ref{fig_phase_diagram} and ref.\cite{Caussarieu_EPL}). We see that the fluctuation amplitude increases when approaching the critical point. 
\begin{figure}[h]  
   \includegraphics[width=0.49\textwidth]{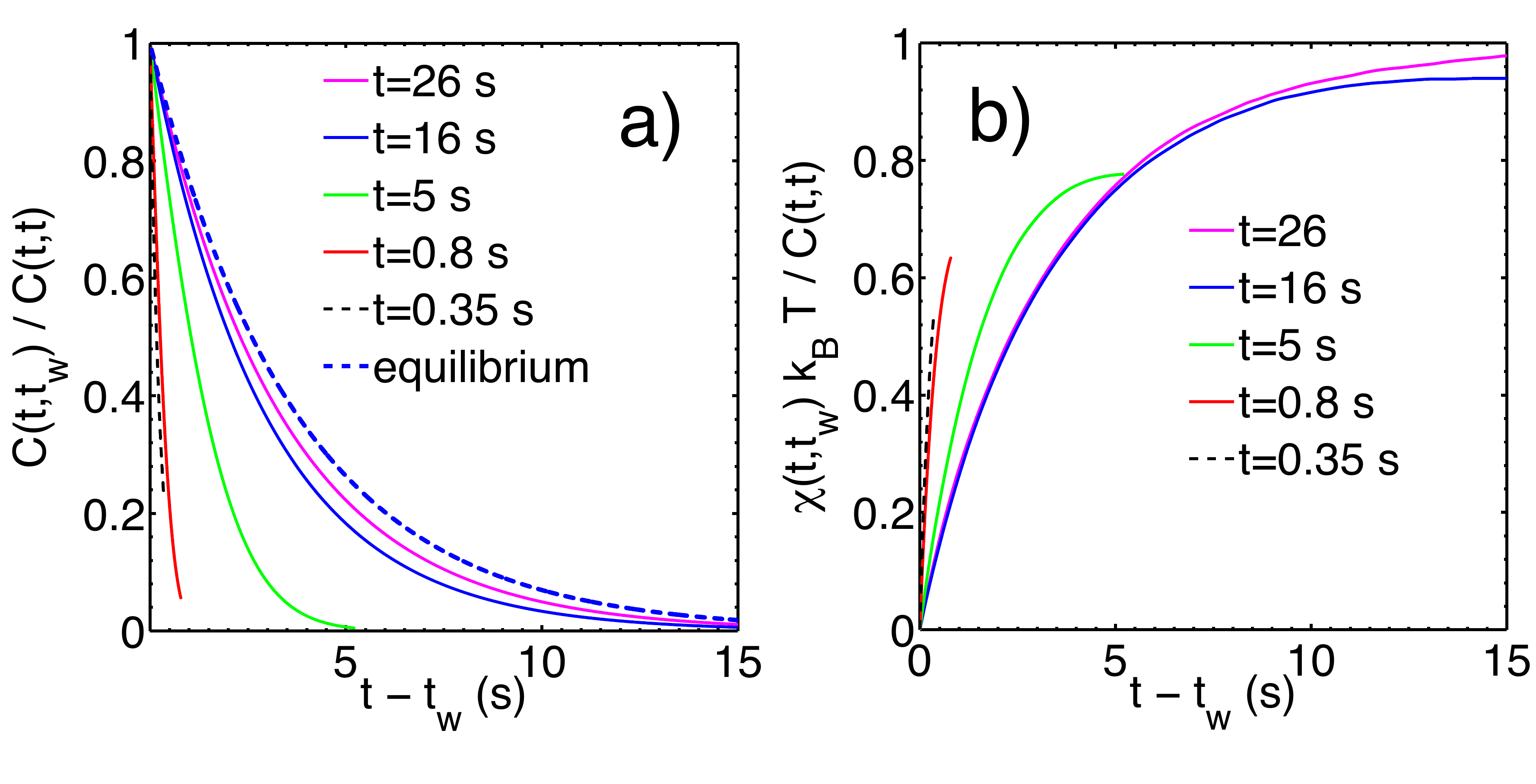}    
     \caption{ The correlations functions  (a) and the integrated  responses  (b) (computed at various fixed times $t$ and $0<t_w<t$ during the relaxation after the quench)  are plotted as a function of 
     $t-t_w$.  
}
 \label{fig_corr_resp} 
\end{figure}
\begin{figure}[h]  
   \includegraphics[width=0.49\textwidth]{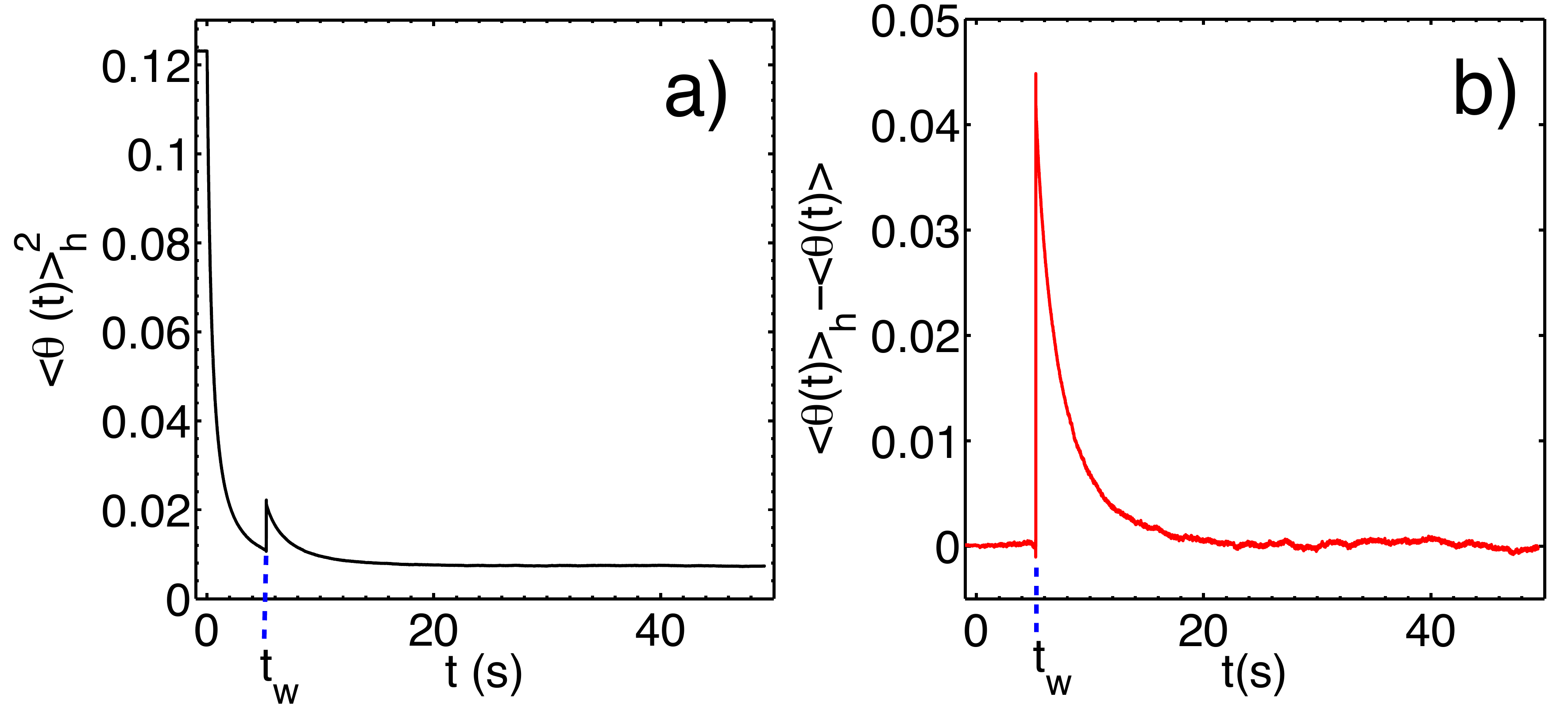}    
     \caption{a) The time evolution has been perturbed at time $t_w$ by a short pulse of amplitude $\delta \epsilon=0.4$ and duration 1ms. 
     b) The response $\Delta \theta$ to the delta perturbation. 
}
 \label{fig_quench_resp} 
\end{figure}

In Fig.\ref{fig_corr_resp}  we plot  $C(t,t_w)=<\delta\theta(t) \delta\theta(t_w)>$ as a function of $t-t_w$ at various $t$ with   $t>tw>0$. We see that $C(t,t_w)$ develops very long decays when $t$ is increased. In order to study FDT we need to measure the response by perturbing the systems 
with a pulse of  amplitude $\delta\epsilon=0.4$ and $\tau_p=1ms$ at time $t_w$. 
As an example, in fig.\ref{fig_quench_resp}a)  we plot the time evolution  perturbed at $t_w=5s$ and in fig.\ref{fig_quench_resp}b) the time evolution of the difference  $<\Delta \theta(t)>=<\theta_\delta(t)>- \theta_o(t)$ between the perturbed $\theta_\delta(t)$ and the unperturbed $\theta_o(t)$.  As it can be seen  in eq.\ref{eq_delta} the amplitude of the perturbation is 
$K \delta\epsilon(t_w) \theta(t_w)\tau_p$. Thus the  impulse response function is $R(t,t_w)=<\Delta \theta(t)>  / (\delta\epsilon(t_w) \theta(t_w)\tau_p)$ for 
$t_w<t$.  We repeat the experiments $N_p$ times by sending at each quench a   pulse at a different time $t_{w,i}$ with $[t_{w,1}=0, .........,t_{w,Np}=20s]$. Then the integrated response  is 
\begin{eqnarray}
\chi(t,t_{w,m})=\sum_{i=m}^{N_t-1}R(t,t_{w,i+1}) ( t_{w,(i+1)} -t_{w,i}) \label{eq_chi} \\
\text{ such that  } \  \  t=t_{w,N_t}  \text{ and  } \chi(t,t)=0 \notag.
\end{eqnarray}
The measured $\chi(t,t_w)$ is plotted as a function of $t-t_w$ for various $t$ in fig.\ref{fig_corr_resp}.b.

\begin{figure}[h!]  
   \includegraphics[width=0.3\textwidth]{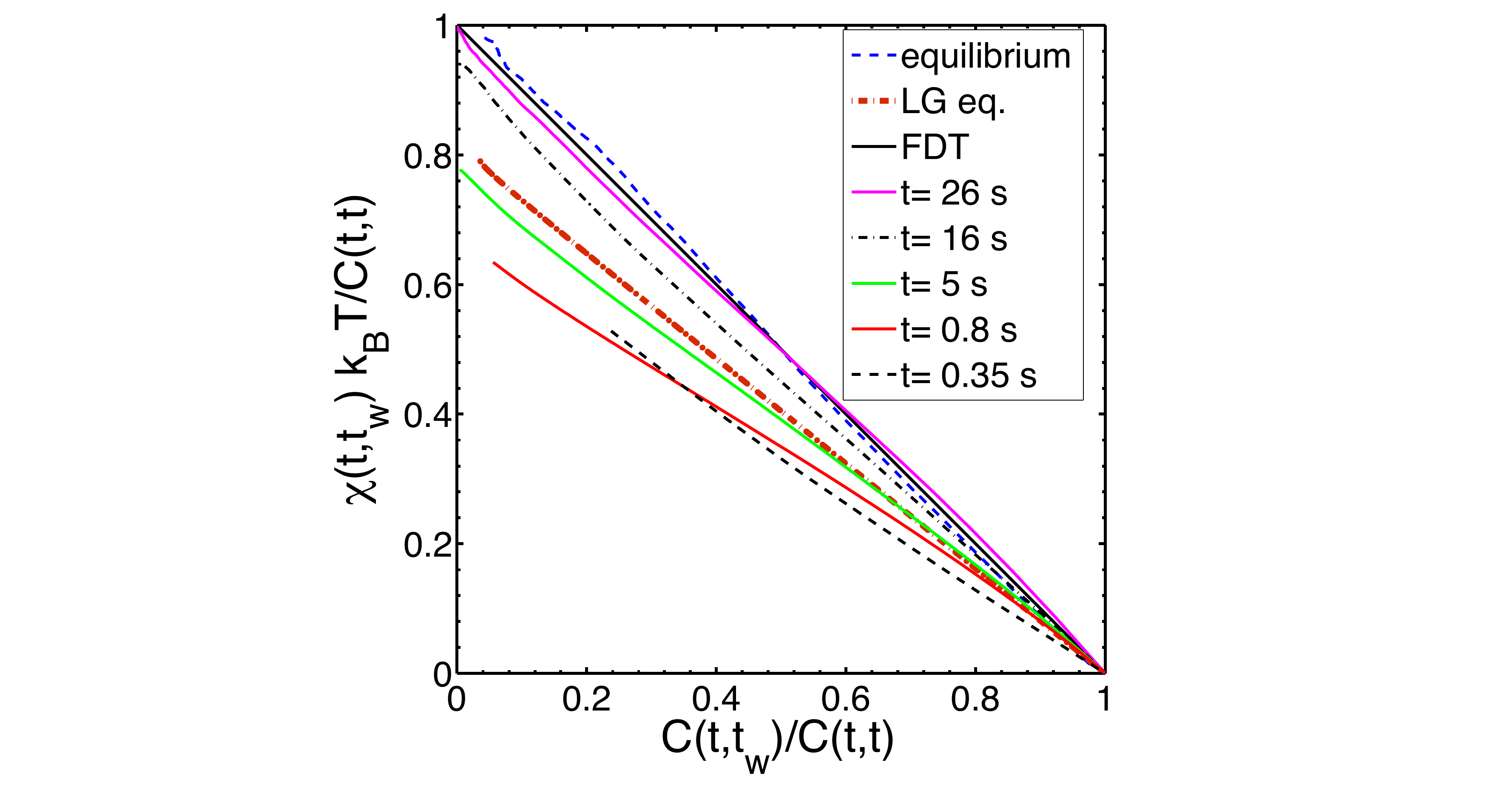}    
     \caption{FDT plot. The function  $\chi(t,t_w)k_BT/C(t,t)$ is plotted as a function of $C(t,t_w)/C(t,t)$ at  various fixed times $t$ and $0<t_w<t$ after the quench. 
     In equilibrium this plot is a the straight line of slope $-1$. We see that at short times $t$ the curves strongly deviate from the equilibrium position. The equilibrium FDT is recovered only for very large $t$.  The red  dashed straight line is the prediction   \cite{Gambassi} for a quench done at $\epsilon=0$ in a Landau-Ginzburg equation.}
\label{fig_FDT} 
 \end{figure}

To check the validity of the standard FDT,  we plot, in fig.\ref{fig_FDT},  $\chi(t,t_w)k_BT/C(t,t)$ as a function  of $C(t,t_w)/C(t,t)$ at various  fixed $t$ with $t_w$ varying in the interval $0\le t_w \le t$. In this plot FDT is a straight line of slope -1. 
We see that for $t$ relatively short, compared to $\tau$, the FDT is not satisfied. In fig. \ref{fig_FDT} we also plot the prediction of ref.\cite{Gambassi} for a quench done at $\epsilon=0$ in a Landau-Ginzburg (LG) equation. We see that for short time the  behavior is quite different from that of the LG equation confirming  that 
the dynamics is not very well described by this equation. The behavior at long time is instead related to the fact that the quench is not performed exactly at $\epsilon_f=0$. 

We now apply the MFDT to these data. In order to do that one has to considers that $\delta \theta$ has a Gaussian distribution whose variance is plotted in fig.\ref{fig_thet_var}. As observable in eq.\ref{eq_MFDTa} we use $O(x(t))=x(t)=\delta \theta$.
 Following the formulation of the MFDT one has to consider the dynamics of  $\Psi(t)$ when a small perturbation $h$ is applied at $t=0$, therefore by  the definitions of $\chi(t,t_w)$ (eq.\ref{eq_chi}) and of $O(t)$,
 we get  $< \delta \theta(t)>_{h}= \chi(t,0) \ h$ because $<\delta \theta(t)>_{0}=0$. Thus at $h\ne  0$ (switched on at $t=0$)  the probability density function for of $\delta \theta$ 
 around the mean is 
\begin{eqnarray}
\pi(\delta \theta(t),h)=\sqrt{1/(2\pi\sigma^2(t))}\exp (-\frac{(\delta \theta- \chi(0,t) \ h)^2}{(2 \sigma_\theta^2(t))}),
\label{eq_MFDT_theta}
\end{eqnarray}
where we assume that if $h$ is small enough then 
the dependence of $\sigma_\theta^2(t)$ on  $h$  can be neglected. 
\footnote{This has been verified experimentally and for the  dynamic described by  eq.\ref{eq_fluct}}. 
 Therefore from the expression of $\pi(\delta \theta(t),h)$, the definition of $\Psi(\delta \theta(t), h)$ and  of $B(t)$ one finds:
$B(t)=\delta \theta(t) \chi(t,0) /\sigma_\theta^2(t)$. Thus eq. \ref{eq_MFDTa} for this particular choice of  variables becomes :
\begin{eqnarray}
{\chi (t,t_w) \over  \chi(t,0)} = -{C(t,t_w) \chi(t_w,0) \over \chi(t,0) \sigma_\theta^2(t_w)}+1
\label{eq_MFDT_theta}
\end{eqnarray}

\begin{figure}[h]  
\begin{center}
   \includegraphics[width=0.3\textwidth]{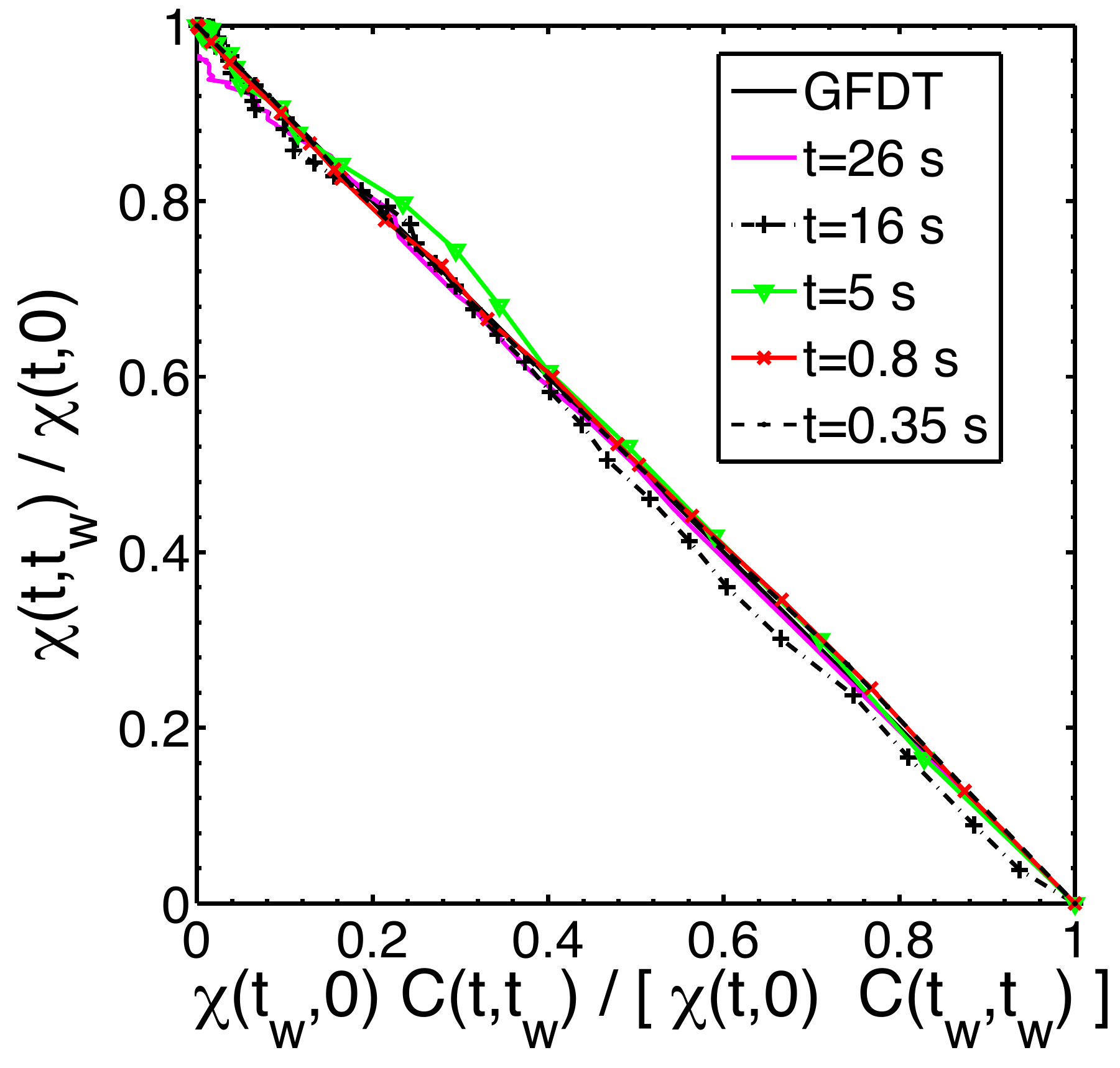}    
     \caption{  In order to verify MFDT  the left  hand side of eq.\ref{eq_MFDT_theta}, i.e. ${\chi (t,t_w) /  \chi(t,0)}$, 
     is plotted   as a function of ${C(t,t_w) \chi(t_w,0) / \chi(t,0) \sigma_\theta^2(t_w)}$. All the data collapse on the straight line of slope $-1$ showing that the MFDT prediction for transient is perfectly verified for any $t$.  Notice that the response and the correlations are the same than those used in fig.\ref{fig_FDT}, but now they have been normalized as prescribed by the MFDT, i.e. eq.\ref{eq_MFDT_theta}. 
}
 \label{fig_MFDT} 
\end{center}
\end{figure}

\begin{figure}[h!] 
\begin{center} 
      \includegraphics[width=0.3\textwidth]{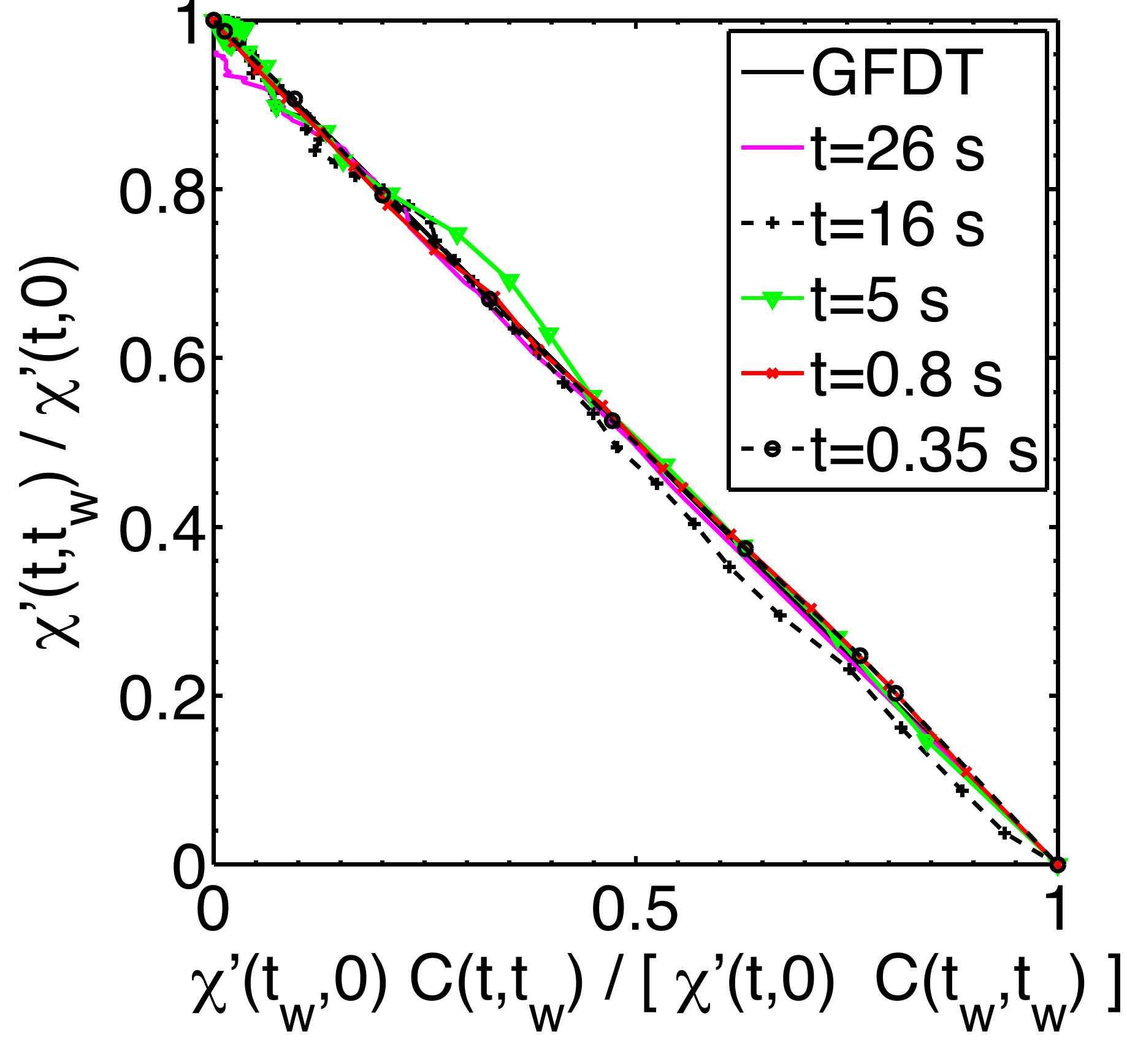}
     \caption{  MFDT recomputed using in eq.\ref{eq_MFDT_theta} the  $\chi'(t,tw)$ defined in the text. The left hand side of eq..\ref{eq_MFDT_theta}  is plotted as a function of the right hand side.  All the data collapse on the straight line of slope $-1$ showing that the MFDT prediction for transient is perfectly verified for any $t$ in this case too. 
}
 \label{fig_GFDT_Rprime} 
\end{center}
\end{figure}

All the quantities in eq.\ref{eq_MFDT_theta} have been already measured. Thus in fig.\ref{fig_MFDT} we plot ${\chi (t,t_w)/\chi(t,0)}$, the left term of eq.\ref{eq_MFDT_theta},  as a function of ${C(t,t_w) \chi(t_w,0) /( \chi(t,0) \sigma_\theta^2(t_w))}$ for various fixed $t$ and $0<t_w<t$. 
We see that all the data points are  aligned on a straight line of slope $-1$ as predicted  by eq.\ref{eq_MFDT_theta}. We clearly see that, in contrast to fig.\ref{fig_FDT} where the standard formulation of FDT is recovered only for very large $t$, MFDT is verified for all times.
As pointed out there is no need in MFDT  to use for $h$ the conjugated variable of $O(t)$. We can use simply  $h=\delta \epsilon$. 
In such a case we define  the response function as  $R'(t,t_w)=<\Delta \theta> /(\delta \epsilon \ \tau_p)$ and the  $\chi'(t,t_w)$ is obtained 
by inserting  $R'(t,t_w)$ in  eq.\ref{eq_chi}. The MFDT computed using in eq.\ref{eq_MFDT_theta} $\chi '(t,t_w)$ instead of $\chi(t,t_w)$   is checked in fig.\ref{fig_GFDT_Rprime}  where the left hand side of eq.\ref{eq_MFDT_theta}  is plotted as a function of the right hand side. We see that the MFDT is verified in this case too .  

All the data presented in this paper correspond to a quench from $\eps_i\simeq 0.25$ to $\eps_f\simeq 0.0$, however the main statistical features, here described,  are independent  on the starting and final points. The final point influences  the duration  of the out of equilibrium state, which  depends on the distance from the critical point. The small difference with the results in ref.\cite{Joubaud_PRL} is due to a slightly non linear response in that reference. 

As a conclusion in this letter  we have applied to a quench a the critical point of Fr\'eedericksz transition (Ftr),  the Modified Fluctuation Dissipation Theorem for transient, proposed in ref\cite{Lacoste_EPL}. We find that although the equilibrium FDT is strongly violated the GFDT is very well satisfied, independently of the chosen response. It is interesting to point out that the result is interesting because although the system presents several differences with respect to the LG equation, it is affected by finite size effects  and the quench is not  performed exactly at the critical point the dynamics still presents features at  the critical quenching. 

We acknowledge useful discussion  with G. Verley and  D. Lacoste. This work has been supported by the ERC contract OUTEFLUCOP.

\end{document}